\documentclass[useAMS,usenatbib]{mn2e}
\usepackage[dvips]{epsfig}
\usepackage{amssymb}

\bibpunct{(}{)}{;}{a}{}{,}
\usepackage{ulem}

\def\dt{\mathrm}

\begin{document}

\title{Numerical counterparts of GRB host galaxies}

\author[Courty, Bj\"ornsson \& Gudmundsson] {S.\ Courty$^{1,2}$, G.\
    Bj\"ornsson~\&~E.~H.~Gudmundsson\\ $^1$Science Institute, University
    of Iceland, Dunhaga~3, IS--107 Reykjavik, Iceland\\ 
    $^2$Laboratoire de l'Univers et de ses Th\'eories, CNRS UMR 8102,\\
    Observatoire de Paris-Meudon,5 place Jules Janssen, 92195 Meudon, France\\
    {\tt e-mail: courty, gulli, einar@raunvis.hi.is} }


\maketitle


\begin{abstract} 

We explore galaxy properties in general and properties of host
galaxies of gamma-ray bursts (GRBs) in particular, using
N-body/Eulerian hydrodynamic simulations and the stellar population
synthesis model, \emph{Starburst99}, to infer observable
properties. We identify simulated galaxies that have optical star
formation rate (SFR) and SFR-to-luminosity ratio similar to those observed
in a well-defined sample of ten host galaxies. Each of the numerical
counterparts are found in catalogs at the same redshifts as the
observed hosts. The counterparts are found to be low-mass galaxies,
with low mass-to-light ratio, recent epoch of formation, and high
ratio between the SFR and the average of the SFR. When compared to the
overall galaxy population, they have colors much bluer than the
high-mass star-forming galaxy population. Although their SFRs span a
range of values, the specific rates of the numerical counterparts are
equal to or higher than the median values estimated at the different
redshifts. We also emphasize the strong relationships between the specific
star formation rate (SFR) and quantities known to reflect the star
formation history of galaxies, i.e.~color and mass-to-light ratio: At
intermediate redshift, the faintest and bluest galaxies are also the
objects with the highest specific rates. These results suggest that
GRB host galaxies are likely to be drawn from the high specific SFR
sub-population of galaxies, rather than the high SFR galaxy
population. Finally, as indicated by our catalogs, in an extended
sample, the majority of GRB host galaxies is expected to have specific
SFRs higher than found in the magnitude-limited sample studied here.

\end{abstract}

\begin{keywords}
cosmology: large-scale structure of Universe 
  -- galaxies: formation -- galaxies: evolution -- gamma rays: bursts
\end{keywords}

\section{Introduction}

Galaxies with a wide variety of properties are observed in the
universe and galaxy sub-populations contribute in a different way to
the overall galaxy properties at different redshifts.  Important
questions in present day cosmology include investigations into how the
different sub-populations can be characterized, how sub-populations
formed at high redshift evolve into galaxies in the local universe and
how these populations contribute to various properties of the overall
galaxy population. In this paper we use a numerical approach to extend
our investigation of galaxy properties \citep[][hereafter referred to
as Paper I]{CBG04} that focused on the specific star formation
rate. While the star formation rate is a 'snapshot' of the stellar
activity in a galaxy, the specific rate is an indicator of how the
galaxy forms its stellar mass relative to the total mass that has been
assembled through its entire lifetime, via mergers and/or
transformation of new accreted gas. The specific rate should therefore
give some insight into the star formation history of galaxies, as do
other properties, such as colors. The specific SFR has been estimated
in a number of observational studies: \cite{Guzman1997} compare the
specific SFRs of compact blue galaxies with other galaxy populations,
\cite{Brinchmann2004} discuss the star-forming galaxies of the SDSS,
\cite{Bell2005} combine infrared data from the \emph{Spitzer Space
Telescope} with optical data, and \cite{Feulner2005} estimate the
specific SFR up to $z=5$.

As in Paper I, we here focus on the properties of a particular
population of galaxies, namely the host galaxies of long-duration
gamma-ray bursts.  The nature of the hosts and their evolution with
redshift are still open questions. The emphasis over the last few
years has been on the host galaxies of long-duration GRBs, which are
seen as a powerful tracer of massive star formation in the
universe. It is now well established that at least some long-duration
GRBs occur at the death of massive stars \citep[e.g.][]{Hjorth2003,
Stanek2003} and in a cosmological context massive stars are very
short-lived. Because of their extreme brightness, long-duration GRBs
are an effective way of locating distant galaxies, most of which are
so faint that they would go undetected in galaxy surveys. In fact,
host galaxies fainter than magnitude 29 have already been detected
\citep{Jaunsen2003}. In addition, GRBs have been detected out to a
redshift higher than $z=6$, and will likely be detected to even higher
redshifts. Rather detailed studies of the local star forming regions
in the hosts is therefore possible (\cite{Berger05a},
\cite{Berger05b}). The most distant burst to date, GRB~050904 at
$z=6.29$ \citep{Tagliaferri2005, Kawai2005}, has already revealed a
number of interesting properties of the interstellar medium of the
host \citep{Totani2005}, while information about galaxy formation and
evolution is yet to be explored in detail, mostly awaiting the
increase of the currently modest sample size (around 50 hosts).

>From individual studies of host galaxies of GRBs \citep{Fruchter99,
  Fynbo03}, comparison of host samples with other sources detected in
various deep surveys \citep{LeFloch2003}, and statistical stellar
population synthesis of the optical and near-infra-red host properties
\citep{Sokolov2001, Chary2002, Christensen04}, indications are that
host galaxies have particular characteristics: These galaxies tend to
be optically sub-luminous, low-mass, blue, star-bursting galaxies,
with young stellar populations, a modest activity of optical star
formation, and perhaps low-metalicity and modest amount of dust
obscuration, although this last feature still needs to be firmly
established.  \cite{LeFloch2003} compare a large sample of host
galaxies of GRBs, observed in the near infra-red, with various galaxy
surveys and find that the observed $K$ and $R-$band magnitudes of the
hosts are comparable to the field sources selected in optical/N-IR
deep surveys, but differ significantly from luminous and dusty
starburst galaxies observed with ISO and SCUBA. Moreover they show
blue $R-K$ colors typical of the faint blue galaxy population in the
field at $z=1$.  Also, SCUBA sub-millimeter observations of GRB host
galaxies performed by \cite{Smith05} suggest that most hosts are not
luminous dusty star-forming galaxies. Furthermore,
\cite{Christensen04} have studied a magnitude-limited sample of hosts
and estimate the ratio between the rest-frame UV star formation rate
and the host optical luminosity. They suggest that the hosts are
similar to those HDF galaxies that have the highest SFR-to-luminosity
ratios.

In Paper I, we showed that among the population still actively forming
stars at low redshift, the high-mass galaxies have much lower specific
SFR than the low-mass galaxies. The non star-forming galaxies, that
span the whole galaxy mass range, are old galaxies while most of the 
stellar populations in the high specific SFR galaxies formed recently. 
At high redshift the trend of increasing specific SFR with
decreasing galaxy mass is also seen, but an interesting point is that
the cosmological evolution is much stronger for the high-mass than the
low-mass galaxies. These trends agree in general well with the
aforementioned observational estimates, to the extent that we
concentrate on the qualitative behavior of the specific SFR. Although
not based on observable properties, the results of Paper I for the
specific SFR suggest that a sub-population of faint and blue galaxies,
some of the characteristics of the GRB hosts, are likely to belong to
the high specific SFR galaxy population, rather than the high-SFR
population.

In this paper, we extend the discussion of the properties of GRB host
galaxies in Paper I, by combining the results of the same simulations
as in that paper with the stellar population synthesis (SPS) code,
\emph{Starburst99} \citep{Vazquez}, to infer observable
properties. SPS codes allow us to compute the spectral energy
distributions of the simulated galaxies at different redshifts. We
identify simulated galaxies that have both similar rest-frame
ultraviolet SFRs and ratios between this SFR and the $B-$band
luminosity as the ten observed hosts of a well-defined sample
(\cite{Christensen04}, hereafter Chr04). The numerical counterparts of
these observed hosts are characterized by a variety of properties,
estimated either directly from the simulation or from the computation
of the SEDs.  In particular, for each galaxy we determine the mass,
the ratio between the SFR and the average of the SFR, $SFR^*/\langle
SFR \rangle$, the specific SFR, the epoch of formation, the $R-K$
color, and the mass-to-light ratio. The properties of the counterparts
are then compared to the overall galaxy population, that is
characterized through the close relationships between the specific SFR
and the color index and mass-to-light ratio. This comparison is,
however, limited by the fact that the observed sample still only
includes 10 hosts, spanning a large redshift range.

Although fairly small, the sample of Chr04 is the only available
homogeneous sample that estimates the ratio between the optical SFR
and the luminosity. Information on other hosts does exist in the
literature and other studies, e.g. \cite{Sollerman05}, use different
star formation rate estimators such as $SFR_{H_{\alpha}}$ and
$SFR_{O_{II}}$, rather than the UV-based indicator adopted in Chr04,
but only for a couple of hosts. A comprehensive study of a large host
sample using all available information awaits future studies. In
Section \ref{sec:hosts} we also briefly discuss the counterparts of
GRB 000911 and 030329, whose $SFR_{UV}$ and $M_B$ values are available
in \cite{Masetti05} and \cite{Gorosabel05}, respectively.

The paper is organized as follows: In
Section~\ref{sec:procedure}, we describe briefly the simulation and
use of the SPS code. Section~\ref{sec:hosts} contains the main 
results of the paper. It starts by discussing the observed host
sample we use (section~\ref{sec:sample}) and the procedure used 
to identify the numerical counterparts (section~\ref{sec:counter}).
We discuss the observational properties of the counterparts in
section \ref{sec:obsprop} and compare their properties with those
of the galaxy population in section~\ref{sec:compare}. 
Section~\ref{sec:disc} concludes the paper.

\section{Numerical procedure}
\label{sec:procedure}

We repeat the simulation used in Paper I in order to obtain galaxy
catalogs at the same redshifts as that of the observed GRB hosts in
the sample of Chr04. We briefly recall the numerical method and we
refer to Paper I for details regarding the simulation. The three
dimensional N-body/hydrodynamical code couples a PM scheme for
computing gravitational forces with an Eulerian approach for solving
the hydrodynamical equations. The dominant processes relevant for
galaxy formation are included: Gravitation, hydrodynamical shocks and
radiative cooling processes. Collisional ionization equilibrium is not
assumed and the cooling rates are explicitly computed from the
evolution of a primordial composition hydrogen-helium plasma. The
cosmological scenario adopted is a $\Lambda-$cold dark matter model
with the following parameters: $H_0=70\ \mathrm{km} \ \mathrm{s}^{-1}
\ \mathrm{Mpc}^{-1}$, $\Omega_K=0$, $\Omega_m=0.3$,
$\Omega_{\Lambda}=0.7$, $\Omega_b=0.02h^{-2}$ with $h=H_0/100$,
$\sigma_8=0.91$. The comoving size of the computational volume is 32
$h^{-1}$Mpc and the simulation has $256^3$ dark matter particles and
an equal number of grid cells. Galaxy formation is introduced using a
phenomenological approach. At each time step, in cells whose gas
satisfies given criteria, a fraction of the gas is turned into a
stellar particle of mass $m_*$ and epoch of formation $a_*$, the
scale-factor corresponding to the cosmic time, $t_*$. The criteria are
the following: The cooling time must be less than the dynamical time,
$t_{\rm cool} < t_{\rm dyn}$; the baryonic density contrast,
$(1+\delta_B)$, must be higher than a threshold $(1+\delta_B)_s=5.5$;
the gas must be in a converging flow, $\nabla \cdot \vec v < 0$; and
the size of the cell must be less than the Jean's length, given by
$\lambda_J = c_s (\pi/G \rho)^{1/2}$. The mass $m_*$ is given by
$m_B(t_0) \Delta t/t_*$, where $\Delta t$ is the timestep, the
characteristic time is taken to be $t_*=\dt{max}(t_{\rm dyn},10^8 \
\dt{yr})$, and $m_B(t_0)$ is the baryonic mass enclosed within the
grid cell. Galaxy-like objects are then defined, at any redshift, by
grouping the stellar particles with a friend-of-friend algorithm. A
simulated galaxy is therefore a collection of stellar particles of
different masses formed at different epochs.  

At any redshift, galaxies are characterized by their mass, $M$,
defined as the sum of the mass of all the stellar particles the galaxy
is composed of; an epoch of formation, defined to be the mass-weighted
average of the epoch of formation of all its stellar particles; and a
star formation rate, $SFR^*$, defined as the amount of stellar
material formed in the previous $10^8$ yr. We also estimate the
specific star formation rate, $\epsilon \equiv \rm{log}(10^{11}$ yr
$SFR^*/M)$, a quantity that measures the efficiency of the conversion
of the gas into stellar material relative to the galaxy mass. We have,
as in Paper I, only considered galaxies with a mass higher than $5
\cdot 10^8$ M$_{\odot}$ in the simulated catalogs. We refer to Paper I
for an illustration of the different sub-populations of galaxies
obtained.

Considering each stellar particle as a homogeneous stellar population,
the total spectral energy distribution (SED) of a galaxy can be
computed from synthesis codes. Evolutionary synthesis codes combine
stellar evolutionary theory to describe the time evolution of model
stars, stellar atmosphere theory to transform quantities from the
theoretical space to the observational one, and the stellar birth
rate, giving the number of stars with a given initial mass formed at a
given time. The stellar birth rate comes from the star formation
history, which gives the number of stars born in a given time, and the
initial mass function (IMF), which gives the relative number of stars
born as a function of mass (see \cite{Cervino} for a discussion
regarding the uncertainties of the synthesis codes). In our procedure,
the star formation history of a galaxy comes directly from the 
formation epoch and mass of all its stellar populations as recorded in 
the simulation. We use the stellar population synthesis model,
\emph{Starburst99} \citep{Leitherer99} in its latest version
\citep{Vazquez}, to derive observable properties. The main change in this
version is the introduction of the Padova stellar evolutionary tracks,
allowing the computation of stellar evolution for old and low-mass
stars as well as high-mass stars.

We start by defining a SED template: we consider a single stellar
population of $10^6$ M$_{\odot}$, assumed to form instantaneously and
evolve passively over a maximal time of 15 Gyr. The SED is computed at
1221 points between 91 {\AA} and 160 $\mu$m. We select in
\emph{Starburst99} the ``Padova AGB'' evolutionary tracks, selection
of the $1992-1994$ Padova tracks with thermally pulsing AGB stars
added, with a metalicity of $Z=0.004=0.2 Z_{\odot}$, where $Z_{\odot}
\simeq 0.02$. We choose a Salpeter IMF, $dP/dm=m^{-2.35}$, with low
and high-mass cut-offs at 0.1 and 100 M$_{\odot}$, respectively. The
other input parameters adopted (supernova cut-off mass, black hole
cut-off mass, wind model, interpolation in mass method, model
atmosphere), are the standard ones. We refer the reader to the
\emph{Starburst99} website\footnote{{\tt
http://www.stsci.edu/science/starburst99/}} for a full explanation.

The total SED of a galaxy at a given redshift is the sum of
the SEDs of all the stellar populations present in the galaxy. The SED
of a stellar particle, at a given redshift, is the SED template
evolved on the time $t(z)-t_*$, in order to take cosmological
expansion into account, and scaled to the mass of the stellar
population with the factor $m_*/(10^6 M_\odot)$. To properly include
the contribution of those stellar particles that formed at the
simulation output time, $t(z)$, the SED is evolved over $10^4$
yr. Note that this procedure is applied to the simulated galaxies
after the simulation run, and not during it. Nevertheless, the
inferred SED of a galaxy accounts for the formation all along the
simulation of different stellar populations. A caveat of the procedure
is that the simulation does not take into account any
evolution of the gas metalicity, and our derived observable properties are
based on a single template. Due to the higher UV-luminosities of
low-metalicity stars, assuming higher metalicity tracks would decrease
the total UV-luminosity. However, in the catalog at $z=1$ almost all of the non
star-forming galaxies have luminosities more than twice the luminosity
computed with solar metalicity tracks, whereas there are only 19
star-forming galaxies with such a large luminosity.

\begin{figure}
\begin{center}
\includegraphics[width=58mm,angle=270]{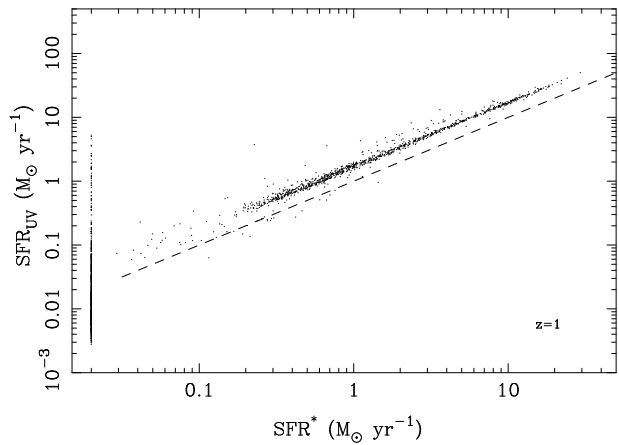}
\caption{Comparison of the star formation rate $SFR^*$ with $SFR_{UV}$
obtained from the synthetic spectrum for galaxies at $z=1$. Non-star
forming galaxies are for display purposes plotted at $SFR^*=0.02$
M$_\odot$ yr$^{-1}$. The diagonal dashed line shows $SFR_{UV}=SFR^*$.}
\label{fig1}
\end{center}
\end{figure}

Star formation rates can be derived from the UV continuum or the
$H\alpha$ luminosity using appropriate proportionality coefficients.
These coefficients are derived from stellar population synthesis
models, involving an initial mass function (IMF) and star formation
history, by looking for a proportionality relation assuming that the
star formation rate is constant in the later phases on given
timescales. These are roughly $10^7$ yr for $H\alpha$ and $10^8$ yr
for the UV. These coefficients therefore depend on the IMF and its
slope, the upper and lower mass cut-off, and the metalicity
\citep{Boselli2001}.  Since one of the purposes of this paper is to
find the numerical counterparts of observed host galaxies, we use the
same indicator of the star formation rate as Chr04, and estimate it
from the UV continuum at 2800 {\AA}, although the SFRs of the two
hosts at the highest redshifts are estimated using the UV continuum at
1500 {\AA} in Chr04. The calibration factor used to convert the
luminosity at 2800 {\AA} into SFR is $1.4\times 10^{-28}$, in units of
(M$_\odot$ yr$^{-1}$)(erg s$^{-1}$ Hz$^{-1}$)$^{-1}$
\citep{Kennicutt98}. In our procedure the luminosity at 2800 {\AA} is
obtained by averaging the spectrum over 20 {\AA} around 2800
{\AA}. Note that, although our spectral energy distributions are
determined at sub-solar metalicity, we adopt the usual calibration to
convert the UV luminosity into SFR that is estimated at solar
metalicity. At lower metalicity the calibration increases due to the
higher luminosity of low-metalicity stars.  The SFR of low-metalicity
galaxies when estimated with a calibration at solar metalicity is thus
overestimated. The evolution of the calibration associated with
different SFR indicators is estimated in \cite{Sullivan2001} and
\cite{Bicker}, for instance. At $Z=0.2 Z_{\odot}$, the latter paper
found that the SFR is overestimated by a factor 1.4, compared to our
calibration.

\begin{figure}
\begin{center}
\includegraphics[width=58mm,angle=270]{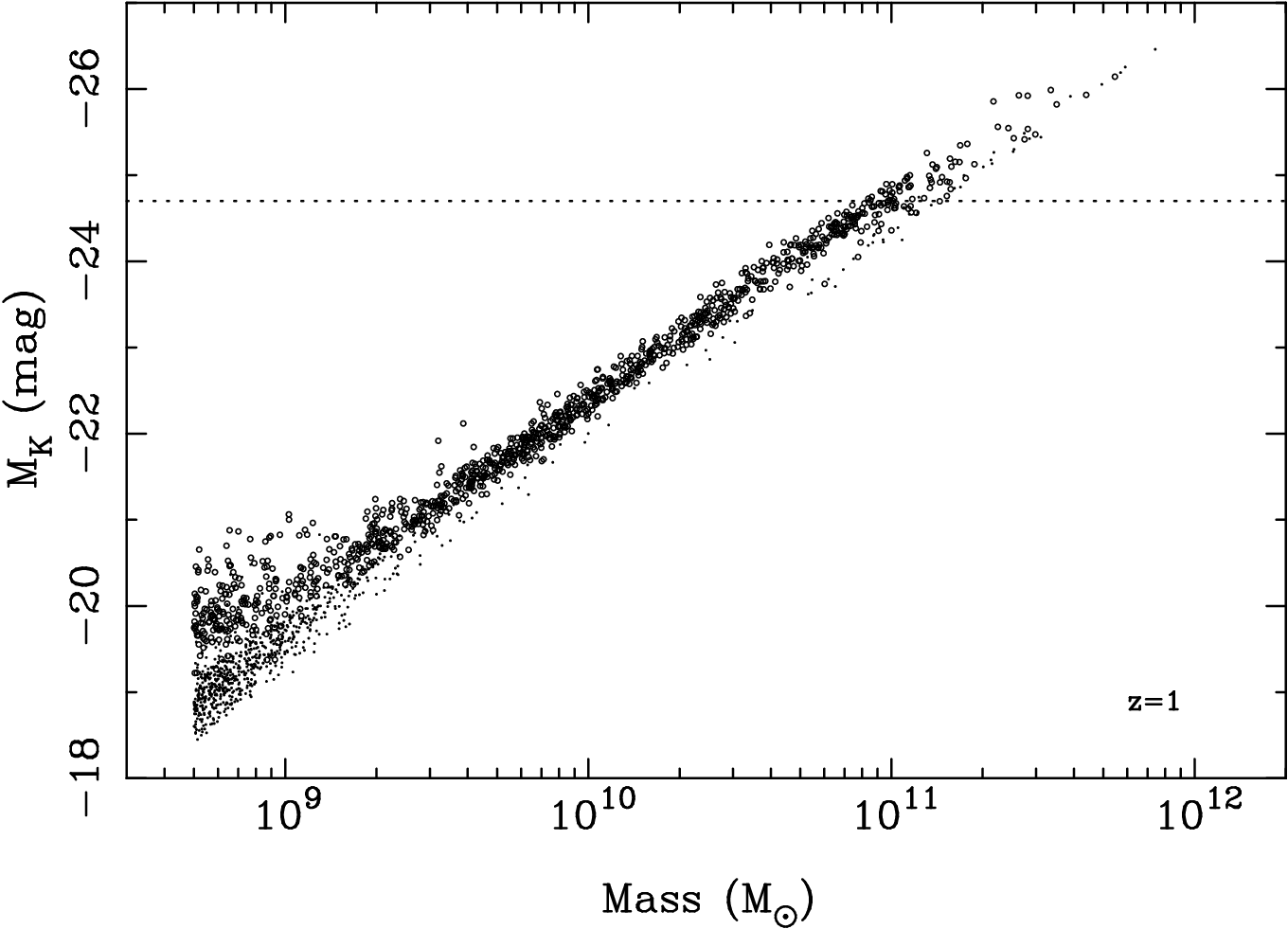}
\includegraphics[width=58mm,angle=270]{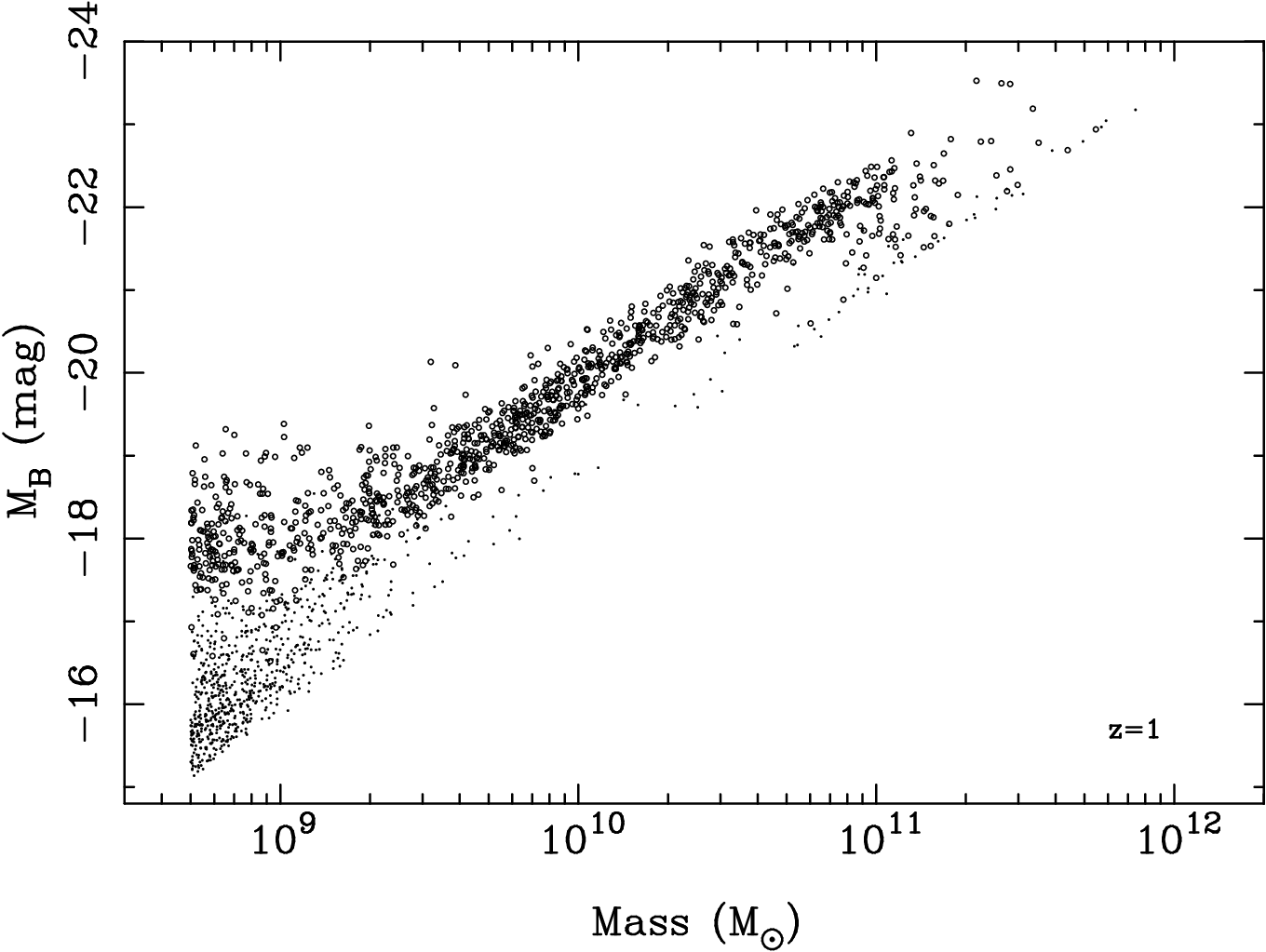}
\caption[]{The mass-magnitude diagrams (upper panel: $K-$band,
lower-panel: $B-$band) for the simulated galaxies at $z=1$.  In both
panels, dots denote star-forming galaxies while the small dots denote
the non star-forming galaxies. The estimate of $M_{K_s}^*=-24.7$
around $z=1$ from the K20 galaxy survey \citep{Pozzetti2003} is marked
by the dotted horizontal line.}
\label{fig2}
\end{center}
\end{figure}

In Figure~\ref{fig1}, we compare the $SFR^*$, computed directly from
the simulation, with the SFR estimated from the synthetic spectrum of
the simulated galaxies at $z=1$. Non-star forming galaxies (i.e.\ with
$SFR^*$=0) are for display purposes plotted at $SFR^*=0.02$ M$_\odot$
yr$^{-1}$. At this redshift, the galaxy population includes 1927
objects, with 1148 star-forming galaxies (as expected, the number of
non star-forming galaxies is larger at $z=0$). The diagonal dashed
line shows $SFR_{UV}=SFR^*$. Despite the scatter in the diagram, it is
clear that the two different estimates are of the same order. The
estimate from the simulations is a factor of about 2 lower than that
obtained from the synthetic spectra.

Figure~\ref{fig2} shows the $B-$ and $K-$band magnitudes as functions
of the galaxy mass for the simulated catalog at $z=1$. The $B$ and
$K-$band magnitudes are computed using the ``Buser's $B3$'' filter and
the ``IR $K$ filter + Palomar 200 IR detectors + atmosphere 0.57''
from the \emph{Galaxev} package \citep{Bruzual2003}. Different symbols
distinguish star-forming and non star-forming galaxies. Note that both
are strongly correlated with the mass. There is much less dispersion
in $M_K$ since the $B-$band luminosities are dominated by the flux
from young massive stars whereas $L_K$ is a better tracer of the
overall stellar population. The corresponding plots at $z=0$ would
show the same trend although quantitatively different. For instance,
the $M_K-$mass relation shows slightly more dispersion and galaxies at
$M_K^*=-24.2$, the characteristic magnitude of the galaxy luminosity
function in \cite{Cole2001}, have a mass of $\sim 10^{11}$
M$_{\odot}$. 

\begin{table*}
\caption{Properties of the numerical counterparts corresponding to the
GRB hosts listed in column one. Following the redshift of the
simulated catalog in col.\ 2, the percentage errors in $SFR_{UV}$ and
the SFR-to-luminosity ratio, $\Delta X$ and $\Delta Y$ defined in the
text, are given in cols.\ 3 and 4. Column 5 shows the mass of the
counterpart, while col.\ 6 shows the epoch of its first formed stellar
population, $t_{form}$ (see Section~\ref{sec:procedure} for
definition) relative to the Hubble time. The mass-to-light ratios in
the $K$ and $B-$bands, respectively, are given in cols.\ 7 and 8,
followed by the corresponding absolute magnitudes. Then comes the
color index, $R-K$, and the next to last column lists the $R(AB)$
apparent magnitude. Finally, a reference number for each host is given
in the last column.}
\label{table1}
\begin{tabular}{@{}lcccccccccccc}
\hline
GRB     &   z  & $\Delta X$ & $\Delta Y$ & M (M$_{\sun}$) & $1-t_{form}/t_{H(z)}$ & $M/L_K$ &  $M/L_B$ &  $M_K$   &    $M_B$   &  $R-K$ & $R(AB)$ & $\#$ \\
\hline
000926 & 2.036 &  0.7 &  0.5 & $1.16 \times 10^{10}$ & 0.76 & 0.36 & 0.36 & $-22.97$ & $-20.83$ &   2.61 &  24.41 & 1 \\
990123 & 1.592 &  5.6 &  3.2 & $2.41 \times 10^{10}$ & 0.81 & 0.47 & 0.59 & $-23.47$ & $-21.08$ &   3.50 &  24.27 & 2 \\
000418 & 1.113 &  5.6 &  8.8 & $8.27 \times 10^{9}$ & 0.86 & 0.35 & 0.32 & $-22.63$ & $-20.57$ &   2.47 &  23.24 & 3 \\
980703 & 0.963 &  1.3 &  0.6 & $3.68 \times 10^{10}$ & 0.88 & 0.52 & 0.66 & $-23.82$ & $-21.41$ &   2.87 &  22.18 & 4 \\
000210 & 0.843 &  6.5 &  5.7 & $5.16 \times 10^{9}$ & 0.89 & 0.47 & 0.54 & $-21.79$ & $-19.50$ &   2.61 &  23.64 & 5 \\
970508 & 0.832 &  3.7 &  0.3 & $7.20 \times 10^{8}$ & 0.89 & 0.30 & 0.25 & $-20.14$ & $-18.18$ &   2.22 &  24.84 & 6 \\
991208 & 0.704 &  1.2 &  0.9 & $1.64 \times 10^{9}$ & 0.90 & 0.41 & 0.45 & $-20.71$ & $-18.46$ &   2.54 &  24.24 & 7 \\
970228 & 0.695 & 10.8 &  0.5 & $1.60 \times 10^{9}$ & 0.90 & 0.49 & 0.56 & $-20.49$ & $-18.19$ &   2.51 &  24.41 & 8 \\
010921 & 0.449 &  1.9 &  1.1 & $8.97 \times 10^{9}$ & 0.92 & 0.68 & 1.01 & $-22.01$ & $-19.42$ &   2.58 &  21.93 & 9 \\
990712 & 0.429 &  3.3 &  2.1 & $1.42 \times 10^{9}$ & 0.92 & 0.38 & 0.36 & $-20.62$ & $-18.55$ &   2.07 &  22.71 & 10 \\
\end{tabular}
\end{table*}

In the following sections we consider other quantities to characterize
galaxies. The lifetime of a galaxy is estimated at a given redshift,
$t(z)-t_{form}$, where $t_{form}$ is the epoch of the formation of the
first stellar population. From this the average of the star formation
rate, $\langle SFR\rangle$, at a given redshift is determined as the
amount of stellar material formed over the lifetime of the galaxy. The
$SFR^*$ is compared with the average through the ratio $SFR^*/\langle
SFR\rangle$, that may be seen as analogous to the so-called birth-rate
parameter, $b$, \citep{Kennicutt94}. The mass-to-light ratios in the
$B$ and $K-$bands are estimated using the solar magnitudes of 5.45 and
3.3, respectively, and the color $R-K$ using the ``Cousins R'' filter
from the \emph{Galaxev} package. Finally, the ratio between the
rest-frame ultraviolet SFR and the luminosity, $SFR_{UV}/(L_B/L^*_B)$,
uses the characteristic magnitude or magnitude at the break of the
local galaxy luminosity function, $M_B^*=-21$. This ratio is sometimes
called the specific SFR as e.g., in Chr04, but we will refer to it by
the term SFR-to-luminosity ratio and reserve the term specific SFR for
the ratio between the SFR and the galaxy mass, as is customary in the
literature.

\section{The numerical counterparts of observed GRB host galaxies}
\label{sec:hosts}

In this section we attempt to identify numerical counterparts to the 
host galaxies of the Ch04 sample, explore their properties, and compare 
the host candidate population with the overall galaxy population. We first 
summarize the relevant properties of the observed sample and our
approach to identifying the numerical counterparts. We then discuss
their observed properties.

\subsection{The observed sample}
\label{sec:sample}

The sample presented in Chr04 is homogeneous and consists of
10 GRB hosts with redshifts between $z=2.037$ and $z=0.433$. The sample 
is magnitude limited ($R<25.3$) and ensures that hosts are bright enough 
to make multi-color photometry possible. The first column in 
Table~\ref{table1} lists the hosts under consideration. The 
observationally determined values of $SFR_{UV}$ and $SFR_{UV}/(L_B/L^*_B)$ 
are given in Table 4 and 5 of Chr04 and we note that the $SFR_{UV}$ entering 
the SFR-to-luminosity ratio is not corrected for internal extinction, 
generally found to be moderate or low. The SFR spans a wide 
range (between 0.8 and 13 M$_\odot$ yr$^{-1}$), but hosts with 
redshifts higher than 0.9 have the highest $SFR_{UV}$, more than 
$\sim 6$ M$_\odot$ yr$^{-1}$, whereas hosts at lower redshifts
have $SFR_{UV} < 2.5 $ M$_\odot$ yr$^{-1}$.

Various observational studies of individual hosts of the Chr04 sample
result in slightly different estimates of the same properties
(\cite{Gorosabel03b} for GRB 000418, \cite{Gorosabel03a} for GRB
000210, \cite{Bloom1999} for GRB 990123, for instance). We have not
attempted to collect all the data existing in the literature, and have
only considered the $SFR_{UV}$ and $SFR_{UV}/(L_B/L^*_B)$ given in
Chr04. All the hosts in Chr04 are also discussed in
\cite{LeFloch2003}, giving either their observed $R$ and absolute
$B-$band magnitudes or their observed $K$ magnitudes and colors,
$R-K$. In addition to the Chr04 sample, we also consider two hosts for
which the SFRs based on the rest-frame ultraviolet flux as well as the
$B-$band magnitudes, are available. We can examine such hosts in a
similar way as for the Chr04 sample. \cite{Masetti05} discuss the
properties of the host galaxy of GRB 000911 at $z=1.06$ and derive a
$B-$band magnitude of around $-18.4$, giving a $L_B/L^*_B=0.09$ with
our $M_B^*$ (see section~\ref{sec:procedure}), and an
extinction-corrected $SFR_{UV}$ of 2.7 M$_\odot$ yr$^{-1}$.  Assuming
an extincted $SFR_{UV}$ to be a factor of 2 lower\footnote{The host of
GRB 991208 has roughly similar $B-$band magnitude and $A_V$ and
has a ratio of un-extincted/extincted SFR within a factor of 2
(Chr04).}, gives a SFR-to-luminosity ratio around 15.  The second host
is that of GRB 030329 at $z=0.168$.  \cite{Gorosabel05} derive a
$SFR_{UV}$ of 0.17 M$_\odot$ yr$^{-1}$ and $M_B=-16.5$, giving a
$SFR_{UV}/(L_B/L^*_B)=10.6$. Note that this value is comparable to the
SFR-to-luminosity ratios derived in the Chr04 sample. The counterparts
of these two hosts are discussed near the end of this section.


\subsection{Counterpart Identification}
\label{sec:counter}

We generated catalogs of galaxies at the same redshifts as the
observed hosts, and looked in each catalog for simulated galaxies that
have both $SFR_{UV}$ and $SFR_{UV}/(L_B/L^*_B)$ nearest to the
corresponding values for each observed host.  We are able to find a
numerical counterpart to each of the 10 observed host of the Chr04
sample, although departures from the observationally inferred values
may be large in some cases. The percentage errors in
Table~\ref{table1} quantify how close from these values the
counterpart is found. They are defined by $\Delta
X=|SFR_{UV}^{obs}-SFR_{UV}^{num}|/SFR_{UV}^{obs}$ and $\Delta
Y=|\epsilon_L^{obs}-\epsilon_L^{num}|/\epsilon_L^{obs}$. In this last
definition, $\epsilon_L$ denotes the SFR-to-luminosity ratio. The
largest departures, when at least one of these quantities is larger
than 5$\%$, are seen for counterparts $\#$3, 5 and 8. The first two
are noted in Chr04 as having the least acceptable fits to their
spectral energy distributions. The host of GRB 000418 ($\#$3) is
extensively discussed in \cite{Gorosabel03b} who analyze its spectral
energy distribution using a variety of synthetic spectral
templates. The adopted SFR does of course depend on the adopted
SED. Also, \cite{Gorosabel03a} find the host galaxy of GRB~000210
($\#$5) to be brighter than the Chr04 estimate: They find that the
$B-$band magnitude is $-20.16$, giving, with their $SFR_{UV}=2.1$
M$_\odot$ yr$^{-1}$, a $SFR_{UV}/(L_B/L^*_B)=4.5$ (10.7 in
Chr04). \footnote{Instead of adopting the nearest counterpart in each
catalog, we could have looked for the second best closest counterparts
or the counterparts within a given error box around each observed
host. Because of the uncertainties in the observationally determined
values and the small number of observed hosts in the sample we
consider here, we do think that searching only for the closest
counterpart is the most reasonable thing to do.}

\begin{figure}
\begin{center}
\includegraphics[width=58mm,angle=270]{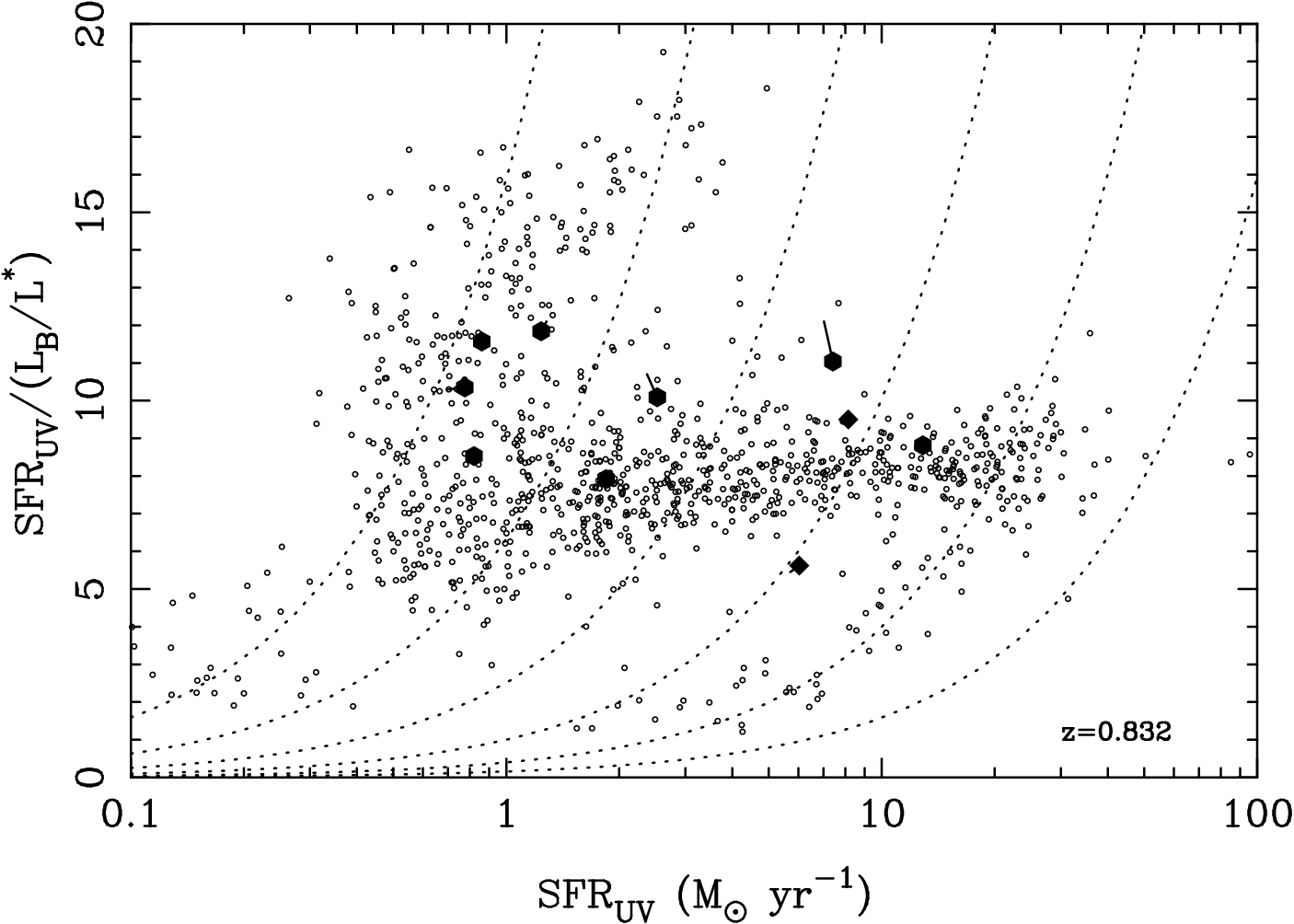}
\includegraphics[width=58mm,angle=270]{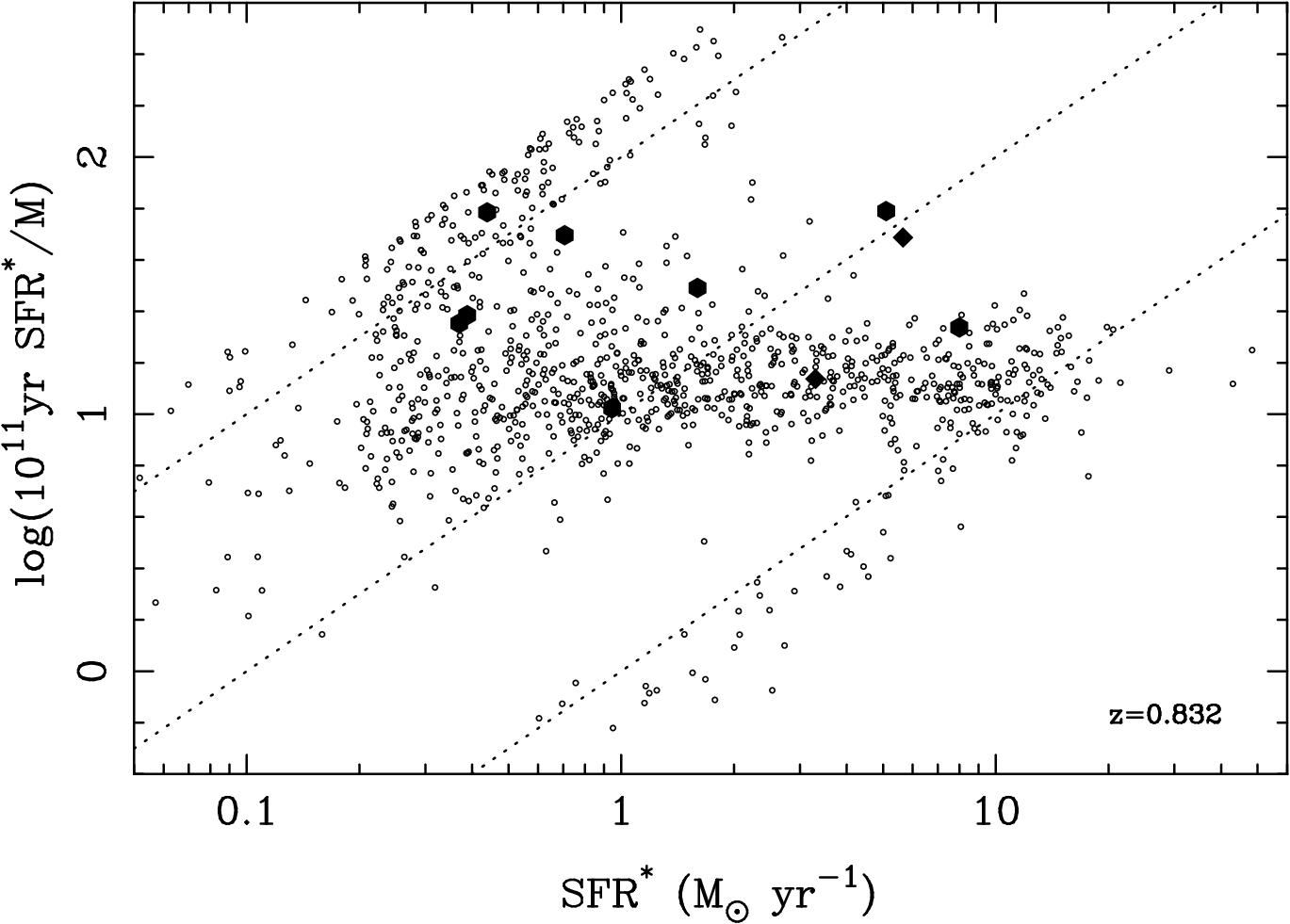}
\caption{SFR-to-luminosity ratio versus $SFR_{UV}$ (upper panel) and
specific SFR versus $SFR^*$ (lower panel) for the star-forming galaxy
population at $z=0.832$ (dots). The dotted curves in the upper panel
denote constant $M_B$-values from $-18$ to $-23$ (left to right). In
the lower panel the dotted lines indicate constant mass of $10^9$,
$10^{10}$ and $10^{11}$ M$_\odot$ (left to right). Although at
different redshifts, the numerical counterparts of the observed host
galaxies (listed in Table~\ref{table1}) are displayed in both panels
(solid symbols): In decreasing order of $SFR_{UV}$, GRB 980703,
000926, 000418, 990123, 000210, 010921, 990712, 970508, 991208, 970228
(the order is the same in the bottom panel with $SFR^*$, except that
the host of GRB 970228 is to the left of GRB 991208). In the upper
panel, for each counterpart a short line segment indicates where the
corresponding observed host galaxy would be found. In some cases the
difference is too small for the line segment to be visible. The two
diamonds denote the counterparts at the highest redshifts.}
\label{fig6}
\end{center}
\end{figure}

Each numerical counterpart is characterized by a number of
properties that are either tabulated in Table~\ref{table1} or
displayed in Fig.~\ref{fig6} and ~\ref{fig8}. The properties
that are estimated directly from the simulation output are the star
formation rate $SFR^*$, the specific SFR, the galaxy mass, the ratio
$SFR^*/\langle SFR \rangle$, the formation epoch of the first stellar
population to form in the galaxy, and the epoch of formation of the
galaxy as defined in section~\ref{sec:procedure}. From the computation
of the SEDs we determine the star formation rate $SFR_{UV}$, the
SFR-to-luminosity ratio, the magnitudes in the $B$ and $K-$bands, the
apparent $R$ magnitude and the color, $R-K$. The mass-to-light ratios
in the $B$ and $K-$bands are then a combination of the primary
simulation outputs and the observable properties.

\subsection{'Observed' properties of the numerical counterparts}
\label{sec:obsprop}

From Table~\ref{table1} we note the following: Low-redshift hosts
($z<0.9$) are low $B-$band luminosity galaxies, $M_B>-20$. We also see
that the counterparts are fainter than $-22$ in the $K-$band at
$z<0.9$.  The 10 counterparts are low-mass galaxies with $M< 4 \times
10^{10}$ M$_{\odot}$ (the highest mass in the simulation at $z=1$,
being $7.4 \times 10^{11}$ M$_{\odot}$). The $B-$band magnitudes and
mass of the counterparts can also be seen in Fig.~\ref{fig6} discussed
below, where in the upper panel the dotted curves denote values of
constant $M_B=-18$ to $-23$ (from left to right) and in the lower
panel the dashed lines indicate values of constant mass of $10^9$ to
$10^{11}M_\odot$ (from left to right). The counterparts have
$L_B/L_B^*$ between 0.074 and 1.46 and $L_K/L_K^*$ between 0.028 and
0.84 (with $M_K^*=-24$), making most of them sub-luminous
galaxies. The mass-to-light ratios in Table~\ref{table1} are
relatively small and more typical of late-type and dwarf galaxies than
large spirals or elliptical galaxies.

The counterparts have apparent $R$-band magnitudes between 22 and 24.4
and comparison between those and the N-IR observational data
reported in \cite{LeFloch2003} for the same hosts (here referred to
as the common sample) shows that the candidate hosts $\#6$ and $\#9$
are the faintest and the brightest, respectively, both of our sample
and of the common sample. The apparent $R$-band magnitudes of
the counterparts are, however, generally slightly brighter than in
\cite{LeFloch2003}. The bluest of the 9 counterparts ($\#10$) for which
the $R-K$ colors of the corresponding hosts were estimated in
\cite{LeFloch2003}, is also the bluest of these 9 observed hosts.
The observed host with the highest color index corresponds the 
counterpart with the second highest color index (after the least 
blue $\#2$). Globally, we do find that 9 counterparts have
$R-K$ colors between 2 and 2.9 and 3.5 for $\#2$. 
\\

In Fig.~\ref{fig6} we superpose the numerical counterparts on a 
plot showing the simulated star-forming galaxy population at $z=0.832$ 
in a diagram of SFR-to-luminosity ratio versus $SFR_{UV}$ (upper panel) 
and in a diagram of specific SFR versus $SFR^*$ (bottom panel). We have 
chosen $z=0.832$ as it is roughly the median redshift of the
sample. Rigorously speaking, only the counterpart of the host at
$z=0.832$ should be compared directly with the simulated catalog, since 
galaxy properties evolve with redshift. It may nevertheless be useful 
to plot the whole sample on a single diagram. We have used different 
symbols (diamonds) to mark the highest two redshifts, and will return to 
the issue of evolution later in this section. Each counterpart in the top
panel is accompanied by a straight line, sometimes smaller than the
symbol, that indicates how well the properties of that numerical
counterpart match those of the observed host (the difficulty in 
obtaining a good counterpart for host $\#3$ is clearly seen). As noted
above, the hosts have various SFRs, be it either the $SFR_{UV}$ or the
$SFR^*$, but they have high SFR-to-luminosity ratio and high
specific rate, as they all lie in the upper half of both panels in
Fig.~\ref{fig6}. The specific SFR and the SFR-to-luminosity ratio
would be even better correlated if the latter quantity is based on 
the $K-$band luminosity, according to the magnitude-mass diagrams 
in Fig.~\ref{fig2}.

\subsection{Comparison with the overall galaxy population}
\label{sec:compare}

Even if some aspects of the simulation may be incomplete, it provides
  us with catalogs of galaxies at various redshifts in which different
  galaxy populations may be distinguished. After defining a particular
  sub-population of galaxies e.g.~similar to the host galaxies of
  GRBs, an important issue is to compare this particular population to
  the overall galaxy population. Such a comparison is in the present
  case limited by the fact that the observed sample includes only 10
  hosts, spanning a wide redshift range where cosmological evolution
  cannot be neglected.

\begin{figure}
\begin{center}
\includegraphics[width=58mm,angle=270]{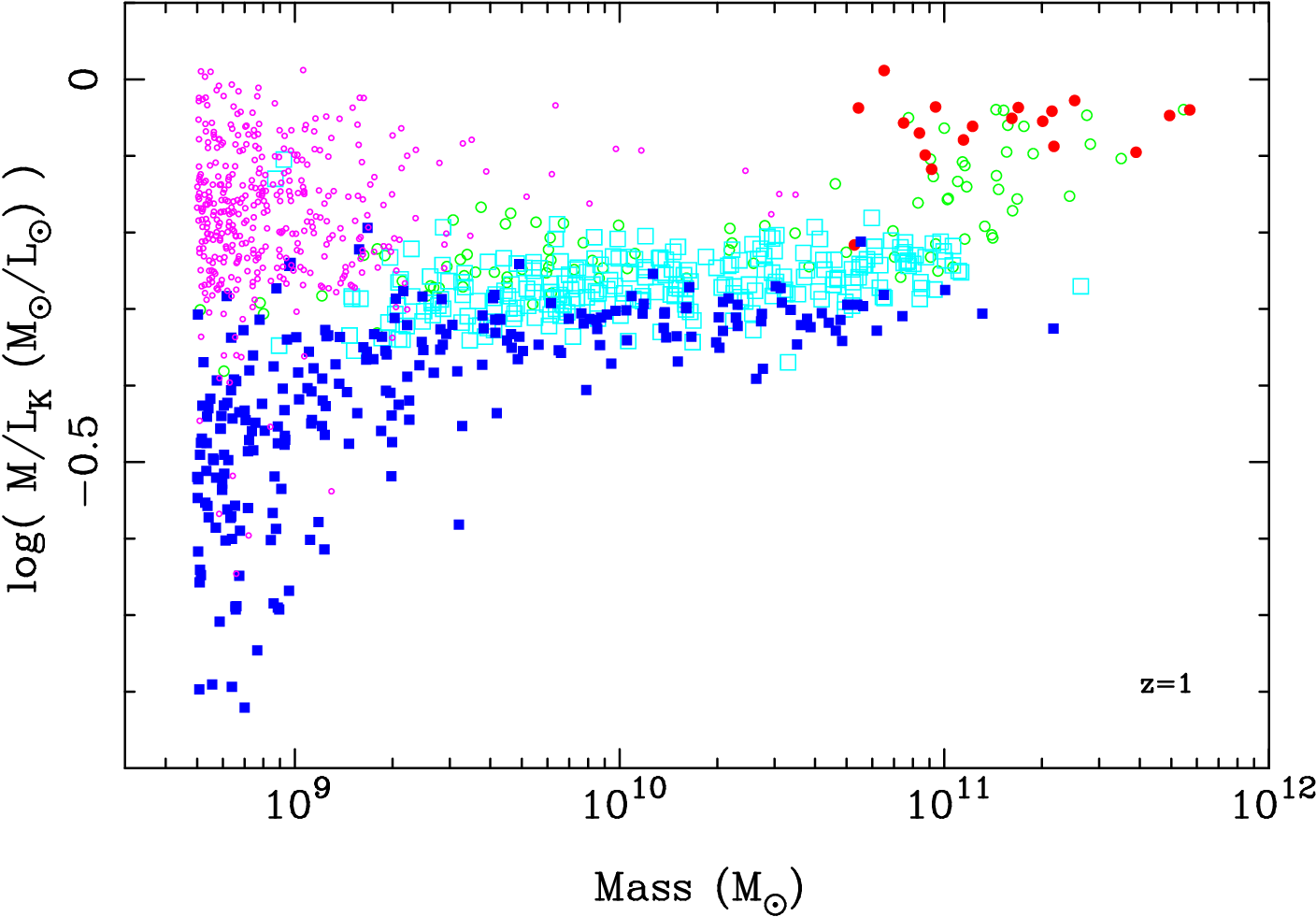}
\includegraphics[width=58mm,angle=270]{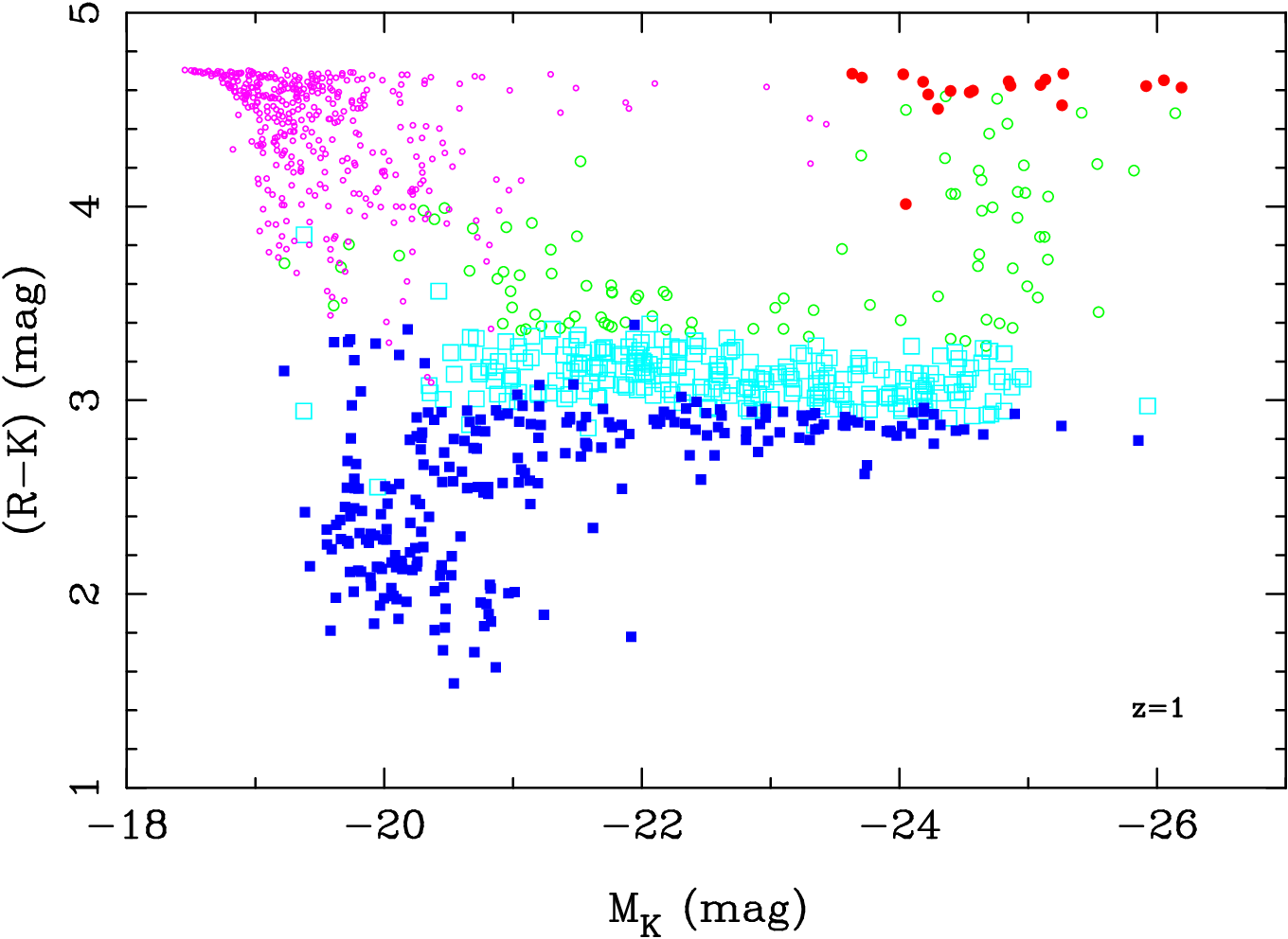}
\caption{Mass-to-light ratio in solar units in the $K-$band as a
function of mass (upper panel) and color-magnitude diagram (bottom
panel) for the simulated galaxy population at $z=1$. For clarity, only
1000 objects randomly selected from the catalog, are plotted. The dots
indicate non star-forming, low-mass ($M<5 \cdot 10^{10}$ M$_\odot$)
galaxies; the filled circles non star-forming, high-mass ($M>5 \cdot
10^{10}$ M$_\odot$) galaxies; the open circles star-forming galaxies
with $\epsilon < 1$; the open squares star-forming galaxies with
$1<\epsilon<1.3$; the filled squares star-forming galaxies with
$\epsilon>1.3$. Galaxies with a mass around $10^{11}$ M$_{\odot}$,
corresponding in Fig.~\ref{fig2} to the characteristic magnitude
$M_{K_s}^*=-24.7$ (or $L_{K_s}^*=1.6 \times 10^{11}$ L$_{\odot}$),
have a mass-to-light ratio of log$(M/L_K) \sim -0.2$.}
\label{fig3}
\end{center}
\end{figure}

\subsubsection{The relationship between the specific SFR and the color index}
\label{sec:sfrcolor}

We first conduct a qualitative comparison between the counterparts and
the overall galaxy population at $z=1$, focusing on the relationship
between the specific SFR and the mass-to-light ratio and color
index. By presenting observable properties at this redshift we
follow-up on and extend the results and discussion of Paper
I. 

Figure~\ref{fig3} presents the mass-to-light ratios and color index of
the whole galaxy population distinguished according to the specific
rate ($\epsilon < 1$, $1<\epsilon<1.3$ and $\epsilon>1.3$). We also
plot the low-mass ($M<5 \cdot 10^{10}$ M$_\odot$) and high-mass
($M>5 \cdot 10^{10}$ M$_\odot$), non-star forming galaxies. We use
the same threshold, $\epsilon=1.3$, as in Paper I. It corresponds
roughly to the peak of the probability density function of the
specific SFR at $z=1$. In addition, galaxies with specific rates below
or above this value contribute about equally to the total star
formation rate at that redshift. The top panel in Fig.~\ref{fig3}
shows the large variation of the mass-to-light ratio with respect to
the galaxy mass, but there is a clear correlation with the specific
SFR (i.e. $\epsilon$). High-mass non star-forming galaxies have large
$M/L_K$, up to about 1, whereas the minimal value for the low-mass
high-specific rate galaxies is $\sim 0.15$. The mass-to-light ratio in
the $B-$band, $M/L_B$, ranges between $\sim$0.09 and $\sim$3, for the
overall population. Recalling the results in Table~\ref{table1}, we
note that the majority of the counterparts have $\log(M/L_K)$ between
$-0.5$ and $-0.3$, typical of our low-mass, high specific SFR
galaxies, although it should be kept in mind that Fig.~\ref{fig3} is
plotted at a single redshift ($z=1$). 

The color-magnitude diagram in the bottom panel in Fig.~\ref{fig3}
shows a large variation of the color index $R-K$, from 1.5 up to 4.7,
consistent with that seen in the recent K20 galaxy survey
\citep{Pozzetti2003}. The high-mass, non star-forming galaxies, that
are also old objects (see Paper~I) have the highest color index,
close to the typical value of old elliptical galaxies at $z \gtrsim
1$. In contrast, the majority of the star-forming galaxies have colors
around 3. The color properties correlate with the specific SFR: The
bluest objects are also faint galaxies with the highest specific
rates. Comparing the data from Table~\ref{table1} with Fig.~\ref{fig3}
shows that the counterparts are clearly bluer than the high-mass
star-forming galaxy population, with color index lower than $\sim$3.
Recalling the correlation between mass and magnitude discussed in the
previous section, we note that the colors and mass-to-light ratios are
also tightly related \citep{Bell2001}. There is also a strong
relation, albeit with some dispersion between mass and SFR, resulting
in intermediate colors of the high-SFR galaxies. Therefore the blue
and faint galaxies, typically characterized by high specific rates,
are not the objects that have the highest SFRs.

Properties like the mass-to-light ratio and color index are generally
considered to provide information on the star formation history of
galaxies or how they assemble their mass. They also tend to be more
tightly correlated with the specific rate than the SFR.  The
$SFR^*/\langle SFR\rangle$, to be further discussed below, is another
property that strongly correlates with the specific rate. As already
pointed out in Paper I, but now confirmed using the calculated
observable properties, we note the consistency between the observed
properties of the GRB host galaxies: If host galaxies are blue and
faint, they are expected not to be high-SFR and early-formed
galaxies. The properties we have found for the counterparts confirm
this conclusion.

\begin{figure}
\begin{center}
\includegraphics[width=58mm,angle=270]{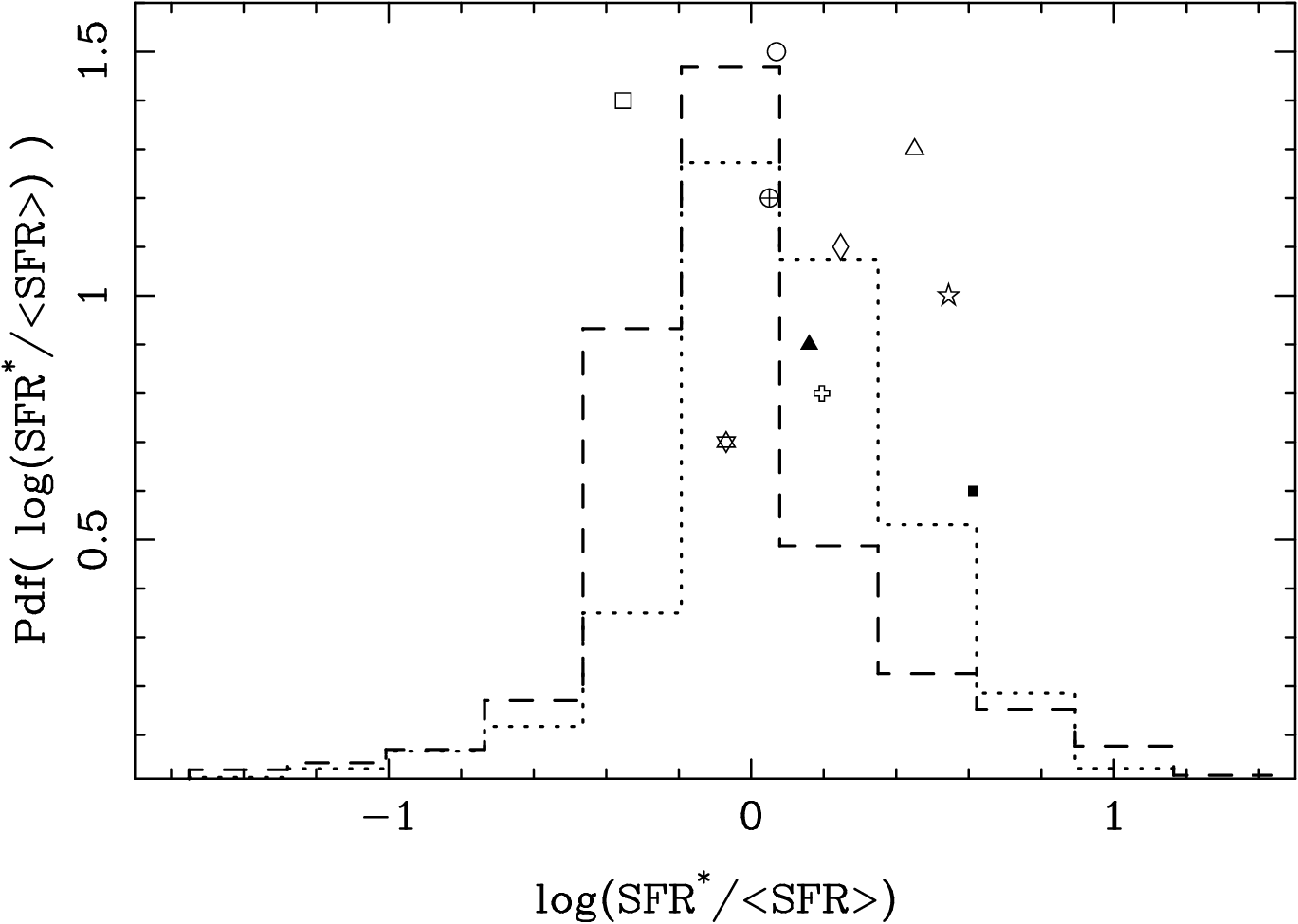} 
\includegraphics[width=58mm,angle=270]{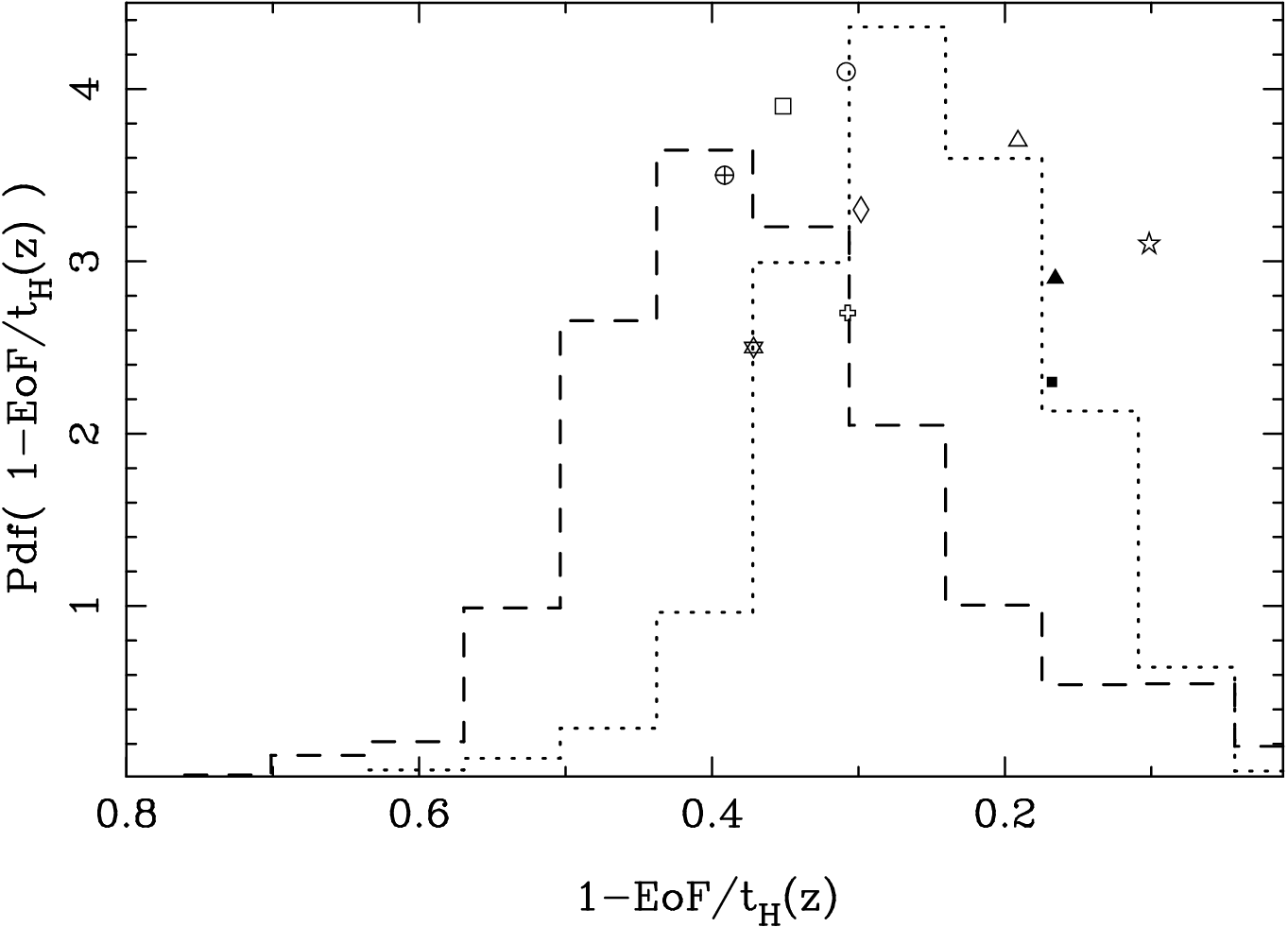}
\caption{Probability density functions of the $SFR^*/\langle SFR
\rangle$ (upper panel) and the epoch of formation $EoF$, when
normalized to the Hubble time (bottom panel), for the star-forming
galaxy populations at $z=2.036$ (dotted) and $z=0.832$ (dashed). The
counterparts of the hosts in Table~\ref{table1} are superposed at
arbitrary ordinate, but in order of decreasing redshift (from top to
bottom).}
\label{fig8}
\end{center}
\end{figure}

\subsubsection{The star-forming activity of the counterparts}

Figure~\ref{fig8} compares the $SFR^*/\langle SFR \rangle$ and the
epoch of formation of the counterparts to that of the star-forming
galaxy population at $z=0.832$ (dashed lines) and $z=2.036$ (dotted
lines), the highest redshift of the host sample. In each panel the
counterparts are superposed on the distributions at arbitrary ordinate
values, but with a decreasing redshift (from top to
bottom). Strictly speaking only two hosts, $\#1$ and $\#6$, should be
compared directly to their respective distributions.

The top panel compares the star-forming activity at a given
redshift of a galaxy to its average activity since the object started
to form its stellar populations. Both distributions peak around unity,
but the high-redshift distribution includes more galaxies with
$SFR^*/\langle SFR \rangle>1$, meaning that high-redshift galaxies are
more active than present-day galaxies. All hosts except $\#2$ have
$SFR^*/\langle SFR \rangle$ close to or above unity (0.85 for
$\#9$). 

The lower panel shows the epoch of formation normalized to the Hubble
time. Comparing the distributions at the two redshifts shows that the
low-redshift one is more extended but includes a non-negligible amount
of objects with recent epochs of formation. All candidate hosts have
ages within 40$\%$ of the age of the universe. We remind the reader
that the distributions are constructed only from the star-forming
galaxies and do not include the old, non star-forming galaxies. The
epoch of formation indicates the epoch at which the galaxy was the
most active and is expected to be quite different from the epoch of
formation of the first stellar populations,
$t_{form}$. Table~\ref{table1} shows that the epoch of formation of
the counterparts are indeed different from $t_{form}$. This particular
time is relatively close to 1 (as defined in Table~\ref{table1}),
meaning that the candidate hosts include an early-formed stellar
population, although not dominant since they are at the same time
young objects.

That the major fraction of the mass of most counterparts was assembled
in the recent times is consistently shown by the specific rate, the
ratio $SFR^*/\langle SFR \rangle$ and the epoch of formation. We note
that the three hosts with the highest $SFR^*/\langle SFR\rangle$
ratios (counterparts $\#$3, 6 and 10), also have the highest specific
rates although their $SFR^*$ and magnitudes differ widely (GRB 000418,
990712 and 970508, see Fig.~\ref{fig6}). They are among the hosts with
the youngest epochs of formation, the bluest $R-K$ colors, and the
lowest mass-to-light ratios (Table~\ref{table1}), pointing again to
the tight correlation between these galaxy properties as discussed
above and to the consistency with the GRB host galaxy properties.

\subsection{Discussion}

\begin{figure}
\begin{center}
\includegraphics[width=58mm,angle=270]{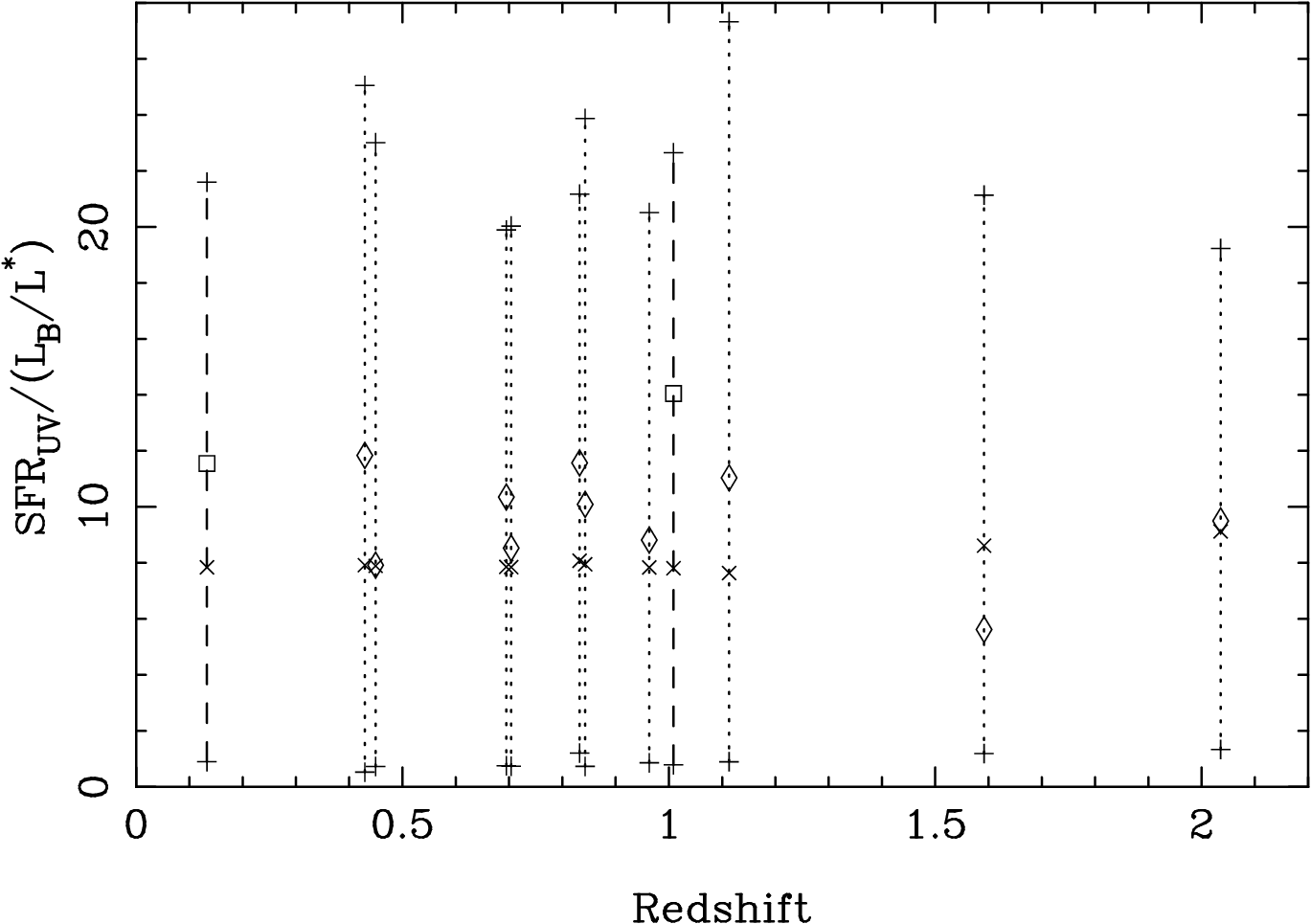}
\includegraphics[width=58mm,angle=270]{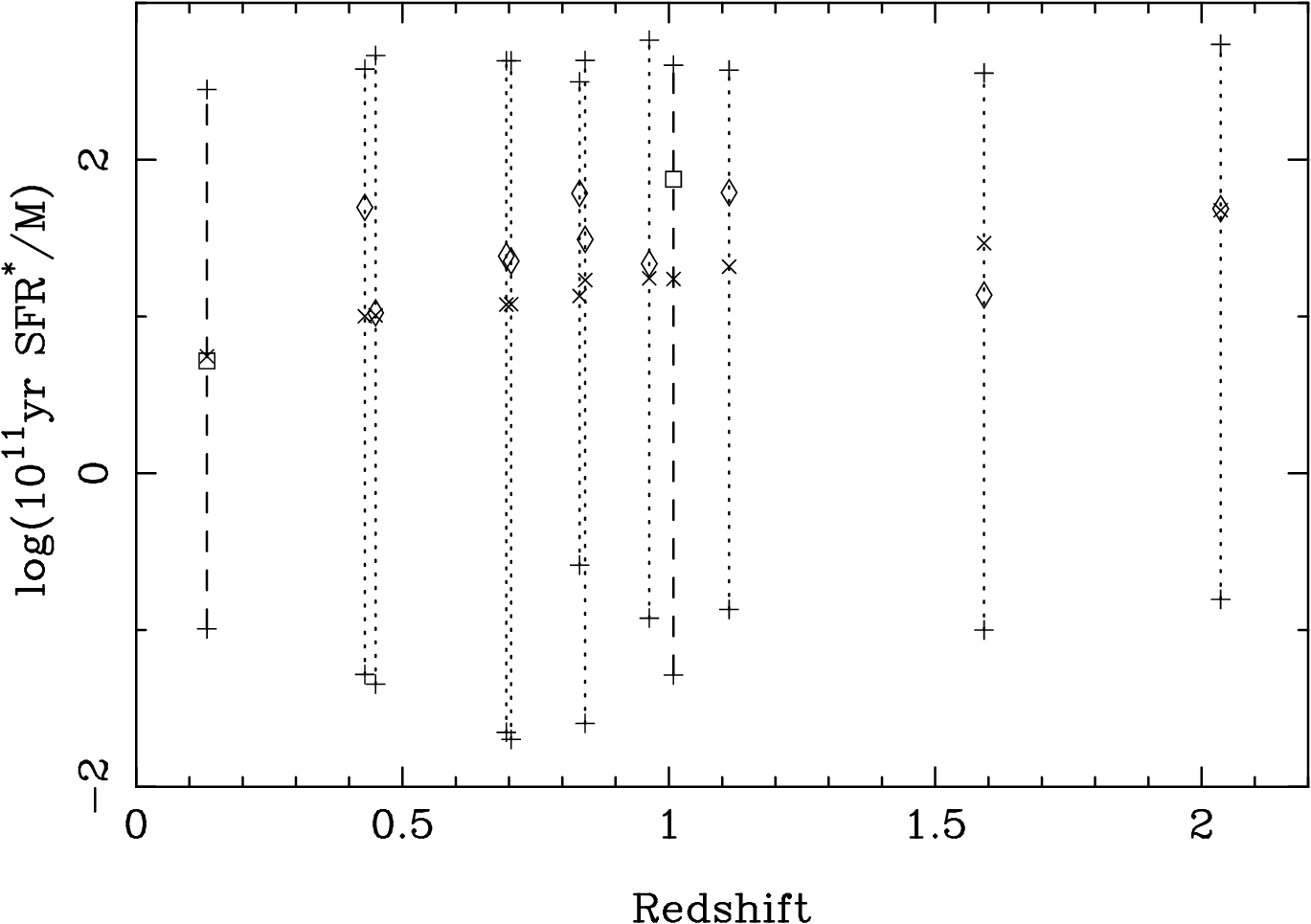}
\caption{The median values (crosses) of the SFR-to-luminosity ratio
(top) and specific SFR (bottom) for the star-forming galaxy
populations at the redshifts of the GRB hosts. The range of values for
each catalog is shown by the length of the vertical lines. Diamonds
refer to the numerical counterpart in each case. The two squares are
the counterparts of GRB 000911 and 030329 hosts that are briefly
discussed at the end of Section~\ref{sec:hosts}.}
\label{fig7}
\end{center}
\end{figure}

Here, we emphasize our main result by comparing the
SFR-to-luminosity ratios and specific SFRs of the numerical
counterparts to the median values of these quantities in the catalogs
they are selected from. These are displayed in Fig.~\ref{fig7}.
The median value of the specific SFR clearly
increases with redshift. Note that at redshifts below $z \sim
0.7$, the minimal specific SFR tends to increase with decreasing
redshift: This is due to the fact that the star formation of massive
galaxies slows down as the redshift decreases and eventually 
ceases in an increasing number of them. They therefore disappear
from the star-forming galaxy population. Comparing the specific SFRs
and the SFR-to-luminosity ratio of the counterparts to the median
values at similar redshift shows that the counterparts in all cases
except one have values higher than the median. The only exception is
the counterpart to GRB~990123 (\#2). The $SFR_{UV}/(L_B/L^*_B)$ of
this particular host is the lowest of the Chr04 sample and its
observed spectral energy distribution was best fit by a star-forming
Sa galaxy type, whereas all of the other hosts were fit by starburst
templates.  The $B-$band magnitude of this host could however be
fainter \citep{Bloom1999}, giving a higher SFR-to-luminosity
ratio. Moreover the $R-K$ color of this host is as blue as most of the
hosts studied in \cite{LeFloch2003}, whereas it is the reddest one of
our sample, as seen above (see Table~1).

It is interesting to note that \cite{Chary2002} estimate the specific
SFRs of hosts $\#$2, 4, 6, 7 and 8, although their observational
study is based on extinction-corrected SFRs, and thus their derived
specific rates are much higher than our results. The observed hosts
$\#$4, 6, 7, and 8 are among those with the highest specific SFR of
the whole \cite{Chary2002} sample. They are also found to have higher
specific rates than local starbursts have. The numerical counterparts
to $\#$6, 7 and 8 have specific rates well above the median values
(the first three objects on the left in Fig.~\ref{fig6}).

The two squares in Fig.~\ref{fig7} show the SFR-to-luminosity ratio
and the specific SFR of the counterparts of the hosts of GRB
000911 and 030329, discussed at the beginning of this section. These
counterparts were searched for in catalogs at slightly different
redshifts from the host redshifts, $z=1$ and $z=0.133$,
respectively. The counterpart of GRB 000911 ($\Delta X=0.3 \%$ and
$\Delta Y=5.2 \%$) is a low-mass, young object ($M=1.1 \times 10^{9}$
M$_{\sun}$, $1-EoF/t_{H(z)}$=0.37) with a high specific rate
($\epsilon$=1.87), blue color $R-K=2.12$ and a mass-to-light ratio
$M/L_K=0.36$. The closest counterpart that we find for the host of GRB
030329 has a lower SFR-to-luminosity ratio than the observed estimate
($\Delta X=1.7 \%$ but $\Delta Y \sim 19 \%$). The candidate host is a
low-mass object ($M=1.4 \times 10^{9}$ M$_{\sun}$), with blue color
$R-K=2.2$, a specific rate of $\epsilon=0.7$ and an epoch of formation
of $1-EoF/t_{H(z)}$=0.45. Compared to the overall population these two
candidate hosts have SFR-to-luminosity ratios and specific SFRs well
above or similar to the median values at the same redshifts.

The fact that the Chr04 sample is magnitude limited and includes
bright hosts may explain why the numerical counterparts are not found
among the objects with the highest possible specific SFRs. The
counterpart of GRB~970508 ($\#$6) is an example. From HST observations
and other studies, this host may be described as a blue compact dwarf
galaxy \citep{Fruchter2000}. Could this host be a prototype of the
general host GRB galaxy population? Interestingly, the counterpart of
this host could be found in the sample of compact galaxies at $z \sim
0.7$ analyzed in \cite{Guzman1997}. The counterpart $\#$6 ($z=0.83$)
with an absolute magnitude $M_B=-18.2$, a mass close to $10^9$
M$_{\odot}$ and a specific rate of $\epsilon \sim 1.8$ falls into the
faintest and highest specific SFR population of compact galaxies in
\cite{Guzman1997} (see their figures 5 and 7). Moreover,
\cite{Sollerman05} discuss three GRB host galaxies and show that they
have similar properties as a sample of compact blue galaxies in the
local universe.

\section{Conclusions}
\label{sec:disc}

In this paper, we have discussed cosmological galaxy properties and
the properties of host galaxies of GRBs in particular, using fully
hydrodynamic simulations of galaxy formation and the stellar
population synthesis (SPS) code, \emph{Starburst99}, to infer
observable properties. An important feature of the numerical procedure
is that the star formation history of galaxies entering the SPS code,
comes directly from the simulation, with each stellar population
contained in a given galaxy treated as a homogeneous population, that
forms instantaneously. We identify objects in the simulation that have
optical star formation rate, $SFR_{UV}$, and SFR-to-luminosity ratio,
$SFR_{UV}/(L_B/L^*_B)$, similar to those estimated in a well-defined
sample of ten observed host galaxies \citep{Christensen04}, the only
available homogeneous sample focusing on this ratio. Each numerical
counterpart is selected from a simulated catalog at the same redshift
as the corresponding observed host, and is characterized by a number
of properties: The $SFR^*$, mass, specific SFR, $SFR^*/\langle SFR
\rangle$, epoch of formation of the first stellar population, and
epoch of formation of the galaxy object. These are obtained directly
from the simulation. In addition, $B$ and $K-$band luminosities,
$R$-band apparent magnitude, $R-K$ color, and mass-to-light ratios are
obtained from the SPS or combination of both results.  It should be
emphasized that some of these properties (e.g.~mass and $SFR^*$) are
estimated directly from the simulations, making their definition
inherently different from those commonly adopted in
observations. 

The \cite{Christensen04} sample includes host galaxies with redshifts
in the range $0.43<z<2.03$. The sample is magnitude-limited
($R<25.3$), with estimated absolute $B-$band magnitudes between
$-21.4$ and $-18.1$. Our counterpart hosts are low-mass galaxies
($M<4 \cdot 10^{10}$ M$_{\odot}$), with low mass-to-light ratios
($M/L_B$ around 0.5); most of them are blue ($R-K<2.9$) and young
galaxies, with epochs of formation (or ages), within 40$\%$ of the age
of the universe at the different redshifts.  Although the $SFR^*$ of
the counterparts varies between $\sim 0.4$ and 8 M$_\odot$ yr$^{-1}$,
the specific SFR is equal to or higher than the median values
estimated for the different catalogs, with the lowest value being
$\epsilon \sim 1$. Because of its strong correlation with the specific
rate, the $SFR^*/\langle SFR \rangle$ also has high values, around
unity or higher. To outline the consistency of such an ensemble
of properties, we compare the counterparts to the overall galaxy
population at intermediate redshift and discuss the strong
relationships between the specific SFR and quantities known to reflect
the star formation history of galaxies, i.e.~color and mass-to-light
ratio. Indeed, the bluest objects are also faint galaxies with the
highest specific rates.

Our identification of simulated galaxies with observed hosts has some
limitations, both from observational and numerical points of view.
Some of the observed hosts may, in other observational studies, be
found to have slightly different SFRs or magnitudes, making their
SFR-to-luminosity ratios uncertain. Although the general agreement
between the simulation and the observations is fairly good, it should
be kept in mind that the moderate resolution and somewhat limited
number of physical processes included in the simulation may affect the
relative number of galaxies in each sub-population. For instance,
these simulations only account for a limited number of
``extreme''-type galaxies, such as extreme starbursts or massive
star-forming galaxies.  Furthermore, we do not include the effects of
dust and, as in \cite{Christensen04}, we only consider extincted
$SFR_{UV}$ in the estimate of the $SFR_{UV}/(L_B/L^*_B)$ ratio. That
may not be a serious drawback since our focus is on low-mass galaxies
that on average suffer less attenuation than the massive ones. In
addition, the amount of dust has been shown to be limited in most host
galaxies of GRBs \citep{LeFloch2006}. The main limitation of this study
is the wide redshift range of the 10 observed host galaxies, with only
a few galaxies in each narrow redshift bin. The inferences made
therefrom need to be confirmed once the sample size has grown by a
factor of 5-10, with a number of hosts at similar redshifts. We should
expect to see still higher specific rates in low-mass objects, as the
simulated catalogs tend to show. The evolutionary effects displayed in
Fig.~\ref{fig7} support this conclusion.

Comparing an expanded host galaxy sample based on {\em Swift} data
with the simulated catalogs and other well-known galaxy
sub-populations, such as compact blue galaxies may be extremely
useful. Hosts galaxies should provide a clearer view into the
formation and evolution of galaxies and the role of different
sub-populations therein. In particular, they will help in
investigating the faint end of the galaxy luminosity function.  Our
results, obtained using a numerical approach, are consistent with and
confirm the picture of GRB host galaxies that has emerged lately: They
tend to have low mass, be blue in color and have relatively high
specific star formation rates. Host galaxies may then belong to the
high specific SFR galaxy population, rather than the high SFR
population. High-resolution simulations are required to determine, in
a more quantitative way, the contribution of the host galaxies to the
overall population, and whether hosts are a part of the average normal
star-forming galaxy population or a sub-population of this one, with
blue and very low-luminosity objects.

\section*{Acknowledgments}
We thank Jens Hjorth and Johan Fynbo for many interesting discussion
sessions on GRB galaxy hosts and their properties. We also thank the
anonymous referee for comments that helped us improve the
presentation. This research was supported in part by a special grant
from the Icelandic Research Council. The numerical simulations used in
this paper were performed on NEC-SX6 of the High Performance Computing
Center Stuttgart (HLRS) of the University of Stuttgart (Germany).

\bibliography{biblio}

\end{document}